\documentclass{article}
\usepackage{arxiv}
\renewcommand{\headeright}{\textit{arXiv} Preprint}
\renewcommand{\undertitle}{\textit{arXiv} Preprint}
\author{Torsten Djurhuus \thanks{The authors are
with the Institute of Physics, Goethe University of Frankfurt am
Main, Max-von- Laue-Strasse 1, 60438, Frankfurt am Main.
(correspondence e-mail:
t.djurhuus@physik.uni-frankfurt.de).} \\
  Goethe-University Frankfurt\\
  \texttt{t.djurhuus@physik.uni-frankfurt.de} \\
  \And
  Viktor Krozer \\
  Goethe-University Frankfurt \& \\
  Ferdinand-Braun-Institut, Leibniz \\
  Institut für Höchstfrequenztechnik\\ 
  \texttt{krozer@physik.uni-frankfurt.de}\\
}

\usepackage[utf8]{inputenc} 
\usepackage[T1]{fontenc}    
\usepackage[hidelinks]{hyperref}       
\usepackage{url}            
\usepackage{booktabs}       
\usepackage{amsfonts}       
\usepackage{nicefrac}       
\usepackage{microtype}      
\usepackage{graphicx}
\usepackage{doi}

\usepackage{amsmath}
\usepackage{amssymb}
\usepackage[mathcal]{euscript}

\usepackage{bm}
\usepackage{epsfig}
\usepackage{wrapfig}
\usepackage{color}
\usepackage{xcolor}
\usepackage{cite}
\usepackage{amsthm}

\usepackage{enumitem}

\usepackage{algorithm}
\usepackage{algpseudocode}

\algnewcommand{\algorithmicand}{\textbf{ and }}
\algnewcommand{\algorithmicor}{\textbf{ or }}
\algnewcommand{\OR}{\algorithmicor}
\algnewcommand{\AND}{\algorithmicand}
\algnewcommand{\var}{\texttt}

\makeatletter
\newcommand{\algrule}[1][.2pt]{\par\vskip.5\baselineskip\hrule height #1\par\vskip.5\baselineskip}
\makeatother

\newcommand{\myFigureScale}{0.725}
\newcommand{\myFigureScaleBig}{0.95}
\newcommand{\myTableScale}{1.1}

\usepackage{hyperref}
\usepackage{cleveref}
\crefname{algorithm}{algorithm.}{algorithms.}

\usepackage{chngcntr}
\usepackage{apptools}

\title{A Novel Phase-Noise Module for 
the QUCS Circuit Simulator. Part II : Noise Analysis.}

\hypersetup{ pdftitle={A Novel Phase-Noise Module for 
the QUCS Circuit Simulator. Part II : Noise Analysis.}, 
pdfsubject={eess.SY, math.CA,
math.DS}, pdfauthor={Torsten Djurhuus, Viktor Krozer},
pdfkeywords={Coupled oscillator phase-noise simulation, noise analysis, 
numerical analysis, circuit simulation, 
circuit analysis, autonomous circuits.}, }

\begin{document}

\maketitle

\begin{abstract}
The paper documents the implementation 
of a novel phase-noise analysis module within 
the open-source QUCS circuit simulator environment. 
The underlying algorithm is based on a rigorous, unified time-domain methodology 
of (coupled) oscillator noise-response, 
recently proposed by the authors. The theoretical approach used to 
develop this model is entirely unconstrained by any 
empirical and/or phenomenological modelling techniques, such 
as \emph{e.g.} LTI and LTV theory, and this 
differentiates it from all prior proposals on this topic. The paper introduces 
important, and previously unpublished, extensions to this framework, 
in the form of novel unified closed-form expressions for both the amplitude and phase-amplitude 
correlation response of a general 
coupled oscillating circuit perturbed by noise. The proposed theoretic framework 
translates into highly robust and efficient numerical algorithms, exhibiting 
several new features and properties which all extend well beyond the scope of 
the noise analysis tools used in current open-source and/or commercial EDAs. The research discussed herein 
has many important scientific and industrial applications \emph{w.r.t.} predicting, synthesizing and optimizing 
the performance of noise-perturbed free-running and coupled autonomous circuits operating under large-signal steady-state conditions. 
These timing circuits are ubiquitous in all modern communication and 
remote-sensing systems and the developed simulation tools will prove to have great 
impact in various areas of industrial circuit design. This paper represents 
second part of a two-part series with the first part 
discussing the implementation of the underlying steady-state analysis module. 
The open-source simulator, discussed and developed herein, 
applies advanced state-of-the-art stochastic modelling techniques, in-order to produce noise simulation tools 
with capabilities and scope which, 
in many areas, exceed what is found in the commercial EDAs currently on the market.
\end{abstract}

\keywords{circuits, phase-noise, circuit simulation, noise analysis, oscillators, coupled 
oscillators, numerical analysis, circuit analysis, autonomous circuits.}

\section{Introduction}
\label{sec0}

Methods for analyzing, synthesizing and optimizing the stochastic response of 
non-linear circuits, operating under large-signal conditions and perturbed by noise, play a critical role 
in the design cycle of circuits with important applications in \emph{e.g.} 
communication and remote-sensing systems. Given the ever-increasing complexity and 
scale of modern circuit schematics, such work necessitates 
using an Electronic-Design-Automation (EDA) simulation program.  
\par
The work herein considers the development and integration of a novel 
phase-noise (PNOISE) tool into the 
Quite Universal Circuit Simulator (\texttt{QUCS}) 
\cite{QUCS_techreport2005,QUCS_report2007}, an open-source distribution published under 
the GNU General Public License (GPL). The resulting extended software package is built as 
a clone of the \texttt{QUCS} \& \texttt{QUCS{-}S} 
\cite{Brinson2015,Brinson2016,Brinson2017} open-source software projects and 
is referred to as the \texttt{QUCS-COPEN} (working title) 
distribution/package/project \emph{etc.} In \cite{Djurhuus2025}, 
the implementation of the \texttt{QUCS-COPEN} Periodic-Steady-State (PSS) 
simulator module was covered. The PNOISE tool, discussed herein, depends directly 
on the numerical output produced by this PSS module. For more details 
\emph{w.r.t.} these topics\footnote{
A very brief recap (for completeness) : the new \texttt{QUCS-COPEN} package 
is constructed from an extended version of the \texttt{QUCS} simulator engine, 
called \texttt{qucsator}, 
which is then linked with a Graphical-User-Interface (GUI) schematic capture application taken 
from the \texttt{QUCS{-}S} distribution and modified to suit the needs of the 
new \texttt{QUCS-COPEN} simulator.\label{sec0:foot1}}, 
the reader is referred to \cite[introduction]{Djurhuus2025}.  
\par
The numerical algorithm behind the \texttt{QUCS-COPEN} PNOISE module, implemented 
herein, is based on a novel theoretical framework known as the Coupled Oscillator Phase-Macro-Model (COSC-PMM), recently proposed 
by the authors in \cite{Djurhuus2022}. The COSC-PMM framework is constructed 
using a novel, rigorous and unified time-domain methodology which involves techniques and ideas 
spanning various branches of mathematics, including differential geometry, topology, 
Floquet theory and nonlinear stochastic integration. It represents a direct 
extension of the established, and highly cited, single/free-running oscillator Phase Macro Model 
(PMM) \cite{Demir2000,Kaertner1990,Traversa2011}. As was documented in \cite{Djurhuus2022}, 
the COSC-PMM includes the single oscillator PMM as a special case, 
and hence not only extends but also \underline{replaces} this previously published model. 
The novel COSC-PMM framework represents the \underline{only} ever published 
Linear-Response (LR) model description of coupled oscillator noise-response which does not rely 
on some form of empirical and/or phenomenological modelling approach, such as \emph{e.g} 
linear-time-invariant (LTI), linear-time-variant (LTV) theory, 
block-diagram models \emph{etc.} All previously proposed models, ever published 
on this topic, can 
hence correctly, and without hyperbole, be characterized as incomplete. A 
prominent example hereof is the Conversion-Matrix (C-MATRIX) methodology where the 
response is calculated using noise-folding/mixing techniques. This frequency-domain 
description models the large-signal oscillating circuit as an internally driven nonlinear mixer 
with the noise signals being fed into the linear RF port. By definition, this 
represents a LTV approach. The C-MATRIX LTV methodology has long been favored 
by a majority of researches in the scientific community as a direct 
approach for developing robust numerical algorithms. 
As a result, most established commercial EDAs, such as \emph{e.g.} 
Keysight-ADS\textsuperscript{\textcopyright} and 
Cadence SpectreRF\textsuperscript{\textcopyright}, all apply this methodology 
in their implementation of large-signal noise analysis tools. 
Despite such proven use-cases it is, nevertheless, 
a well-established fact \cite{Demir2000,Kaertner1990,Suarez2009} that 
these LTV C-MATRIX models, applied to 
(coupled) oscillating systems perturbed by noise, are unable to fully capture 
the underlying stochastic response; a fact which manifests in an 
artificial singularity at the carrier (zero offset) in the calculated spectrum \cite{Demir2000,Kaertner1990,Suarez2009}. 
As this singularity is approached, numerical noise will accumulate in 
the calculated solution rendering the output of the algorithm dubious and 
unreliable. 
As such, this issue not only invalidates the calculated spectrum 
directly at the singularity, but also corrupts the solution 
in some open range of this point\cite{Suarez2009}. 
These types of fundamental and structural problems remain an 
inescapable realty when dealing with any empirical or 
phenomenological model as the underlying theoretic description, by definition, 
must be incomplete and lacking. 
\par
As the \underline{only complete}, non-phenomenological/empirical LR model of coupled oscillator noise-response, 
the COSC-PMM is not tainted by any of the problems or issues described above. 
The produced spectrum is correctly calculated at all offsets from the carrier, 
no artificial singularities are observed in the output and no 
numerical noise issues invalidates the computed result 
or plays any role in the performance of the method.
The QUCS-COPEN package is the first ever, to our knowledge, 
EDA project to include a PNOISE module based on the novel COSC-PMM methodology. 
Consequently, it is thus also the \underline{first ever} 
EDA project to include a PNOISE analysis module, 
applicable to coupled circuits, which is correct, in the strictest sense, meaning that it does 
not rely on any kind of LTV, LTI, block-diagram or other phenomenological/empirical 
modelling approach to calculate the spectrum.

\subsection{A Note on Alternative Open-Source Engines : \texttt{qucsatorRF}, \texttt{GnuCap}, 
\texttt{ngspice} \& \texttt{Xyce}.}
\label{sec0:sub0}

The \texttt{QUCS-S} \cite{Brinson2015,Brinson2016,Brinson2017} software package 
ships with its own \texttt{qucsator} engine clone called \texttt{qucsatorRF} (see \cref{sec0:foot1}) 
and both the \texttt{QUCS} \cite{QUCS_techreport2005,QUCS_report2007} and 
\texttt{QUCS-S} packages 
link to other non-\texttt{QUCS}, \texttt{SPICE} based, open-source simulator engines, 
including : \texttt{GnuCap} \cite{GnuCapManual}, 
\texttt{ngspice} \cite{ngspiceManual} and \texttt{Xyce} \cite{XyceManual2025}. 
All these external engines were discussed in 
\cite{Djurhuus2025} and the reader should consult this paper for details.
From \cite{Djurhuus2025}, only the \texttt{ngspice} packages supported any type 
of PSS analysis of autonomous simulation. Likewise, the 
\texttt{ngspice} manual also contains the only mention of large-signal 
noise analysis of autonomous circuits. Quoting the \texttt{ngspice} manual 
\cite[sec. 1.2.8]{ngspiceManual} : "Results of PSS are the
basis of periodical large-signal analysis like PAC or PNoise". As was the 
case with the \texttt{ngspice} PSS module this noise analysis implementation 
is labelled \emph{experimental code} (see 
discussion in \cite{Djurhuus2025}). Furthermore, neither the 
PAC or PNoise modules, as far as we can tell, are mentioned anywhere else in the \textasciitilde750 page 
manual. Given these facts, one must conclude that these features have 
not yet been implemented or, at the very least, have not yet been properly tested or documented. 
Thus, a review of the current state of the available open-source simulator engines, 
\emph{w.r.t.} available PNOISE tools, leads to only one conclusion  : no credible 
alternative options exist for analysis of large-signal noise-response of autonomous circuits (clocks/oscillators), 
not to mention coupled circuits, in any of the major open-source 
simulator engines discussed here.

\subsection{Outline of paper}

Below, various theoretical and practical aspects \emph{w.r.t.} the 
COSC-PMM methodology and the implementation of the 
\texttt{QUCS-COPEN} PNOISE analysis tool, are discussed. 
\Cref{sec1,sec1:sub1,sec1:sub2} 
reviews the theory behind the COSC-PMM framework and 
the corresponding implemented numerical algorithm. In \Cref{sec1:sub3} the 
original COSC-PMM scheme, proposed in \cite{Djurhuus2022}, 
which was limited to the phase-noise response, is extended to include 
expressions for both the amplitude and 
cross-correlation noise response components. These important additions 
are published herein for the first time and will serve to further improve 
the scope of the COSC-PMM model framework. \Cref{sec1:sub4,sec1:sub5} 
cover the various practical issues involved \emph{w.r.t.} the actual 
implementation \& integration of the PNOISE tool into the \texttt{QUCS-COPEN} package\footnote{
The text in \cref{sec1:sub4,sec1:sub5} will, in places, contain brief discussion of 
C++ code \cite{Stroustrup2013} (the \texttt{QUCS} program language) 
including the use of some C++ notation and keywords.}. 
In \cref{sec2}, the newly implemented \texttt{QUCS-COPEN} noise analysis module 
will be demonstrated on a 
varied set of example oscillator circuits, both coupled and free-running. 
Then in \cref{sec2:sub1}, the calculated results are validated through a comparison study with the commercial Keysight-ADS\textsuperscript{\textcopyright} 
simulator. Finally, \cref{sec3} summarizes 
the obtained results and outlines future work \& projects, currently in the pipeline \emph{w.r.t.} 
to the \texttt{QUCS-COPEN} project.

\section{Background Theory \& PNOISE Module Implementation.}
\label{sec1}

The noise-perturbed circuit equations (KCL/KVL) are written using 
a standard Modified-Nodal-Analysis (MNA) notation \cite{Demir1998}\footnote{
The noise-free version of \cref{sec1:eq1} was analyzed in part-I \cite{Djurhuus2025} of 
this series. Hence, brief passages of text in this section, involving steady-state 
(\emph{i.e.} noise-free, PSS) topics, 
may constitute a repeat (to some degree) of the discussion in \cite{Djurhuus2025}. 
This approach was chosen in the interest of completion and readability.}

\begin{equation}
\dot{q}(x(t)) + i(x(t)) + s(t) + B(x(t))\xi(t) = 0
\label{sec1:eq1}
\end{equation}

where, $x(t) : \mathbb{R} \to \mathbb{R}^n$, represents the $n$-dimensional circuit state, 
vectors $q(x),i(x) : \mathbb{R}^n \to \mathbb{R}^n$, hold the reactive and resistive (KCL/KVL) contributions, 
respectively, $s(t) : \mathbb{R} \to \mathbb{R}^n$ contain the contributions of 
all independent sources in the circuit, $B(\cdot) : \mathbb{R}^n \to \mathbb{R}^{n\times p}$ 
is the \emph{noise modulation matrix} \cite{Traversa2011,Kaertner1990,Demir2000} and $\xi(t) : \mathbb{R} \to \mathbb{R}^p$ 
holds $p$ uncorrelated, zero-mean, unit-power white-noise sources.
\par
The topic of discussion here is noise-analysis of autonomous 
circuits operating under large-signal conditions, which implies the existence 
of a special asymptotic periodic solution to (noise-free) \cref{sec1:eq1}, 
$x_s(t) : \mathbb{R} \to \mathbb{R}^n$, the so-called Periodic Steady State (PSS). 
This solution is assumed stable and periodic, $x_s(t_0 + T_0) -x(t_0) = 0$, 
where $t_0,T_0 \in \mathbb{R}_+$ are parameters 
representing some asymptotic time offset and the steady-state period, 
respectively. The limit-cycle, $\gamma$, 
represents the unique 1-dimensional $\omega$-limit set (attractor)
traced out by the asymptotic PSS solution. Thus, $x_s(\eta) \in \gamma$, 
for all $\eta$ and, $\lim_{t\to \infty}\Vert x(t)-\gamma \Vert = 0$, for any solution $x(t)$ to \cref{sec1:eq1} 
inside the stable manifold (\emph{i.e.} the basin of attraction) of $\gamma$. 
The paper \cite{Djurhuus2025} (part-I of series) 
contains a review of the various proposed methods for 
deriving this solution numerically, 
including a detailed description of the specific methodology used to implement the \texttt{QUCS-COPEN} 
PSS analysis module.

\subsection{Deriving the Floquet Decomposition.}
\label{sec1:sub1}

Given the PSS, $x_s(t) : \mathbb{R}\to \mathbb{R}^n$, 
(see \cite{Djurhuus2025} for details), the circuit equations in \cref{sec1:eq1}
can be linearized around this solution. The resulting 
MNA Linear-Response (LR) equations have the form

\begin{equation}
d(C(t)\delta x(t))/ dt + G(t)\delta x(t) = 0 
\label{sec1:eq2}
\end{equation}

where, $\delta x(t) : \mathbb{R} \to \mathbb{R}^n$, is the LR state vector and 
 $C(t) {=} dq(s)/ds,G(t) {=} di(s)/ds|_{s=x_s(t)} : 
\mathbb{R} \to \mathbb{R}^{n\times n}$, are two time-dependent, periodic 
Jacobian matrices governing the linearized dynamics, with $i,q$ defined above in \cref{sec1:eq1}. The PSS is assumed stable and 
a solution to \cref{sec1:eq2} is then calculated 
by integrating these differential equations forward in time. 
As the system in \cref{sec1:eq2} is linear this solution can be expressed 
in-terms of the system State-Transition-Matrix (STM) $\Phi(t,s) : \mathbb{R}\times \mathbb{R} 
\to \mathbb{R}^{n\times n}$. Furthermore, with the time-dependence being periodic, 
this STM can be decomposed using Floquet theory \cite{Demir1998}

\begin{equation}
\Phi(t,s) = \sum_{i=1}^n u_i(t) \exp(\mu_i[t-s])v_i^{\top}(s)C(s)
\label{sec1:eq3}
\end{equation}

where $\mu_i \in \mathbb{C}$ is the $i$th \emph{Floquet characteristic exponent}, 
$u_i(t),v_i(t) : \mathbb{R} \to \mathbb{C}^n$ are the $i$th, $T_0$ periodic, Floquet and dual Floquet vectors, 
respectively, and the $T_0$ periodic Jacobian matrix, 
$C(t) : \mathbb{R} \to \mathbb{R}^{n\times n}$, was defined above in \cref{sec1:eq2}. From \cref{sec1:eq3}, the Monodromy Matrix (MM), 
$\Phi_{\text{\tiny M}} \in \mathbb{R}^{n\times n}$, is derived as 

\begin{equation}
\Phi_{\text{\tiny M}} \equiv  \Phi(T_0,0) = \sum_{i=1}^n \lambda_i u_i(T_0) v_i^{\top}(0)C(0)
\label{sec1:eq4}
\end{equation}

where $\lambda_i = \exp(\mu_i T_0)$ is the $i$th \emph{Floquet characteristic multiplier}.  
Given periodicity, $u_i(0) = u_i(T_0)$, and the orthogonality condition, 
$v_i^{\top}(t)C(t)u_j(t) = \delta_{ij}$ \cite{Demir1998}, it follows from inspection 
of \cref{sec1:eq4} that $u_i(0)$ is an eigenvector of the MM 
with eigenvalue $\lambda_i$, and in this way, the entire set, $\{u_i(0)\}_{i=1}^n$, 
is derived. Using this set of eigenvectors as initial conditions, the system in \cref{sec1:eq2} is 
integrated forward in time to produce 
the complete set of time-periodic Floquet vectors $\{u_i(t)\}_{i=1}^n$ \cite{Kaertner1990,Demir2000,Djurhuus2009}. 
This set spans the so-called tangent-bundle\footnote{
A tangent-space, $\mathbb{T}_{x}\mathbb{M}$, on a $k$-dimensional sub-manifold, $\mathbb{M}$, 
of $\mathbb{R}^n$, at the point $x$, is an affine copy of vector-space $\mathbb{R}^k$ with origin 
at $x$. A tangent-bundle along an embedded set, $\gamma \in \mathbb{M}$, then simply refers 
to the disjoint union of these tangent-spaces along this base-space \emph{i.e.}
$\mathbb{T}_{\gamma}\mathbb{M} = \cup_{x\in \gamma} \mathbb{T}_{x}\mathbb{M}$. 
This description also hold for the special case $\mathbb{M} = \mathbb{R}^n$ 
(\emph{i.e.} no sub-manifold) Likewise, a dual (co-variant) tangent-space, $\mathbb{T}_{x}^*\mathbb{M}$, 
on the $k$-dimensional sub-manifold, $\mathbb{M}$, of $\mathbb{R}^n$, 
at the point $x$, is an affine copy of dual vector-space $\mathbb{R}^{k*}$ with origin 
at $x$. Equivalently, a dual tangent-bundle is defined as
$\mathbb{T}_{\gamma}^*\mathbb{M} = \cup_{x\in \gamma} \mathbb{T}_{x}^*\mathbb{M}$ 
which again also holds for the special case $\mathbb{M} = \mathbb{R}^n$.\label{sec1:foot1}} 
along the limit-cycle base-space \cite{Djurhuus2009,Djurhuus2022}

\begin{equation}
\mathbb{T}_{\gamma}\mathbb{M} = \text{span}\{ u_1(t), u_2(t), \cdots, u_n(t)\}
\label{sec1:eq5}
\end{equation}
 
The adjoint LR equations, corresponding to the system in \cref{sec1:eq1}, are 
written \cite{Demir1998}

\begin{equation}
C^{\top}(t)d(\delta x(t))/ dt - G^{\top}(t)\delta x(t) = 0 
\label{sec1:eq6}
\end{equation}

A solution to \cref{sec1:eq6}, for a given initial condition, 
can be found by integrating this system backwards 
in time. This solution is then be expressed in-terms of the adjoint STM

\begin{equation}
\Psi(t,s) = \sum_{i=1}^n v_i(t) \exp(-\mu_i[t-s])u_i^{\top}(s)C^{\top}(s)
\label{sec1:eq7}
\end{equation}

and the corresponding adjoint MM is given as

\begin{equation}
\Psi_{\text{\tiny M}} \equiv \Psi(-T_0,0) = \sum_{i=1}^n \lambda_i v_i(-T_0) u_i^{\top}(0)C^{\top}(0)
\label{sec1:eq8}
\end{equation}

Repeating the arguments introduced above, $v_i(0)$ is an eigenvector of the adjoint 
MM in \cref{sec1:eq8} with eigenvalue $\lambda_i$ and the dual set, 
$\{v_i(t)\}_{i=1}^n$, is calculated by integrating \cref{sec1:eq6}, with 
this initial eigenvector-set, backwards in time. The 
dual Floquet vectors span the dual tangent bundle (see \cref{sec1:foot1}) 
along the limit-cycle base-space

\begin{equation}
\mathbb{T}_{\gamma}^*\mathbb{M} = \text{span}\{ v_1(t), v_2(t), \cdots, v_n(t)\}
\label{sec1:eq9}
\end{equation}


\subsubsection{The Concept of a Floquet Mode \& Floquet Mode Response.}
\label{sec1:sub1:sub1}

Herein the Floquet vector function, $u_i(t)$, 
is often referred to as the $i$th Floquet mode. Sometimes, 
the entire triplet of operators $\{u_i(t),v_i(t),\mu_i\}$ is 
also referred to as such. The $i$th 
\emph{Floquet mode response}, $z_i(t) : \mathbb{R} \to \mathbb{C}^n$, 
then represents the projected noise-driven LR dynamics along 
the $i$th Floquet mode, $z_i(t) = \kappa_i(t)u_i(t)$, 
where $\kappa_i : \mathbb{R} \to \mathbb{C}$ is the first-order noise-response along this 
vector. From \cref{sec1:eq1,sec1:eq3} it follows that

\begin{equation}
\kappa_i(t) = \int_{-\infty}^t \exp(\mu_i[\eta - t])v_i^{\top}(\eta)C(\eta)B(x_s(\eta))\xi(\eta) d\eta
\label{sec1:eq10}
\end{equation}

with noise operators, $B,\xi$, defined in \cref{sec1:eq1}. Importantly, we 
will often also refer to the mode response, $z_i(t)$, as the $i$th mode 
for simplicity. This should not cause any confusion.

\subsubsection{The Special PSS Floquet Mode.}
\label{sec1:sub1:sub2}

It is a well-established fact \cite{Kaertner1990,Demir2000,Traversa2011,Djurhuus2009} 
that autonomous circuits, with a stable PSS, will produce a special single real 
Floquet mode with zero characteristic exponent. This special operator, 
traditionally represented by the first mode $\{u_1,v_1,\mu_1=0\}$, 
is special and unique as it is the only mode, assuming an asymptotically stable PSS, that has a Floquet characteristic exponent 
equal to zero $\mu_1 = 0 \Rightarrow \lambda_1 = 1$. It describes the LR dynamics projected onto the 
tangent-direction of the limit-cycle, $\gamma$, with $u_1(t) \propto \dot{x}_s(t)$ for all $t$. 
The dynamics in this direction are neutrally stable, \emph{i.e.} no retracting force, 
which is characterized mathematically by a characteristic exponent with real part equal to zero. Let, 
$\alpha(t) : \mathbb{R} \to \mathbb{R}$, describe the phase offset along $\gamma$ 
caused by the noise perturbations projected onto this set. From \cref{sec1:eq10}, 
one can derive an expression for the phase increment

\begin{equation}
\alpha(t+\tau) - \alpha(t) = \int_{t}^{t+\tau} v_i^{\top}(\eta)C(\eta)B(x_s(\eta))\xi(\eta) d\eta
\label{sec1:eq10a}
\end{equation}

The increment in \cref{sec1:eq10a} is bounded, however, it should be clear that the 
phase variable, $\alpha(t)$, itself is not. In fact, it is a well established result 
\cite{Kaertner1990,Demir2000,Traversa2011,Djurhuus2009} that, asymptotically with time, 
this variable becomes normally distributed with power $\langle(\alpha(t) - \kappa)^2\rangle = ct$ where the 
positive scalar, $c \in \mathbb{R}_{+}$, is known as the \emph{phase diffusion constant}, and $\kappa$ represents 
the asymptotic mean value.

\subsection{The Novel COSC-PMM Modelling Framework.}

\label{sec1:sub2}

The model considers $k$ free-running, asymptotically stable autonomous oscillator units
coupled through some type of network. This circuit topology is referred to 
as a \emph{$k$-ensemble}. It is assumed that the ensemble reaches a $T_0 \in \mathbb{R}_+$ periodic 
synchronized state, $x_s(t_0) = x_s(t_0+T_0)$, \emph{i.e.} 
the ensemble PSS. The concept of a unique \emph{phase-manifold}, $\mathcal{M}$, 
corresponding to a given $k$-ensemble configuration, was first discussed 
in the paper \cite{Djurhuus2022}, recently published by the authors. 
It refers to a closed, $k$-dimensional, smooth sub-manifold embedded 
in $n$-dimensional state-space, $\mathbb{R}^n$, and parameterized by a set of coordinates 
indexing the phase-state of the coupled circuit. 
Using various established results and ideas from topology and differential geometry, 
the existence of this unique manifold was proven for a synchronized circuit 
in \cite{Djurhuus2022}. 
The analysis further showed how the ensemble 
limit-cycle, $\gamma$, constitutes a 1-dimensional 
sub-manifold embedded in this space, \emph{i.e.} $\gamma \hookrightarrow \mathcal{M}$. Let $\mathbb{T}_{\gamma}\mathbb{M}$ represent 
the tangent-bundle (see \cref{sec1:eq5,sec1:foot1}) along $\gamma$ 
on $\mathbb{M}$. A standard result of COSC-PMM theory, introduced in \cite{Djurhuus2022}, then 
predicts the existence of exactly $k$ unique Floquet modes 
spanning this bundle. As the numbering of Floquet modes is 
arbitrary we are free to choose these to be the first $k$ vectors 
in the set

\begin{equation}
\mathbb{T}_{\gamma}\mathbb{M} = \text{span}\{u_1(t),u_2(t),\cdots,u_k(t)\}
\label{sec1:eq11}
\end{equation}

with the remaining set of $n{-}k$ Floquet vectors spanning the 
amplitude bundle $\mathbb{T}_{\gamma}\mathcal{I}$ \cite{Djurhuus2022}. 
From \cref{sec1:eq5}, the tangent-bundle along $\gamma$ 
is then decomposed into phase and amplitude sub-bundles as

\begin{equation}
\mathbb{T}_{\gamma}\mathbb{R} = \mathbb{T}_{\gamma}\mathcal{M} \oplus \mathbb{T}_{\gamma}\mathcal{I}
\label{sec1:eq12}
\end{equation}

Let $\{\phi_i(t)\}_{i=1}^k$ denote the 
$k$ Floquet modes\footnote{Note, that strictly speaking, 
operators $\phi_i(t)$ \& $\psi_i(t)$ represent \emph{mode responses} 
characterizing noise driven phase/amplitude LR 
along the $i$th phase/amplitude Floquet mode 
(\emph{i.e.} corresponding Floquet vectors, see \cref{sec1:sub1:sub1}). However, as explained in \cref{sec1:sub1:sub1}, 
for simplicity, these operators will also be referred to herein as \emph{modes}. \label{sec1:foot4}} 
on $\mathbb{T}_{\gamma}\mathcal{M}$, which govern the 
LR noise-driven phase dynamics on $\mathcal{M}$; \emph{i.e.} the \emph{phase-modes}. 
Likewise, $\{\psi_i(t)\}_{i=1}^{n{-}k}$ then refers to the 
set of $n{-}k$ \emph{amplitude-modes} (see \cref{sec1:foot4}) governing the LR noise amplitude dynamics in directions not 
tangential to $\mathcal{M}$.  Referring to \cref{sec1:sub1:sub2}, 
the first phase-mode $\{u_1,v_1,\mu_1=0\}$, 
requires special attention. As the response, $\alpha(t) : \mathbb{R} \to \mathbb{R}$ grows without 
bound this response cannot be linearized and must be included in the total phase of the 
various operators. For this purpose the effective time-variable $\iota(t) : \mathbb{R} \to \mathbb{R}$ 
is introduced 

\begin{equation}
\iota(t) = t + \alpha(t) \label{sec1:eq12x}
\end{equation}

The remaining response is collected in the residual phase and amplitude 
actions $\Phi_{\delta}, \Psi_{\delta} : \mathbb{R} \to \mathbb{R}^n$,

\begin{equation}
\Phi_{\delta}(\iota(t)) = \sum_{i=2}^{k} \phi_i(\iota(t)), \quad 
A_{\delta}(\iota(t)) = \sum_{i=1}^{n{-}k} \psi_i(\iota(t))  \label{sec1:eq14} 
\end{equation}

where the residual phase LR action is seen to exclude the 
special first mode (sum index starts at $i=2$) for the exact 
reasons mentioned in \cref{sec1:sub1:sub2} and above in connection with \cref{sec1:eq12x}. 
Given this setup, an expression for the $k$-ensemble LR, 
on the phase-manifold $\mathcal{M}$, and tangential to this space, can be written as

\begin{equation}
x_{\text{LR}}(t)|_{\mathcal{M},\text{\tiny tangential}}  = x_{\text{LR}}(t)|_{\mathbb{T}_{\gamma}\mathcal{M}} = x_s(\iota(t)) + \Phi_{\delta}(\iota(t)) 
\label{sec1:eq15}
\end{equation}

Since the above relation is restricted to direction tangential $\mathcal{M}$, 
it describes the \emph{phase-noise response} of the coupled circuit.
From \cref{sec1:eq15}, an expression for the asymptotic 
system correlation matrix on $\mathcal{M}$, 
$C_s(\tau) : \mathbb{R} \to \mathbb{C}^{n\times n}$, can be produced. 
The corresponding power spectral density, $\mathfrak{L}^{(\nu)} : \mathbb{R} \to \mathbb{R}$, 
\emph{i.e.} the PNOISE spectrum, around the $\nu$th harmonic of the PSS carrier, 
then follows from Fourier transforming the $\nu$th harmonic envelope of $C_s(\tau)$. 
The procedure described here was carried out in 
\cite[sec. 6]{Djurhuus2022} and following some rather lengthy and involved calculations, 
the following expression, slightly re-written, was derived

\begin{equation}
\mathfrak{L}^{(\nu)}(\omega_m^{(\nu)}) = \frac{ (1 - a^{(\nu)} )(\omega_0^2\nu^2c)
-2b^{(\nu)} \omega_m }{ (0.5\omega_0^2\nu^2c)^2 + (\omega_m^{(\nu)})^2} +
\sum_{\rho} \sum_{l=1}^{k-1} \frac{ \Upsilon^{(\nu)}_{l\rho} \bigl[
2|\mu_{l+1,r}| + (\omega_0^2\rho^2c) \bigr] +
2\Delta^{(\nu)}_{l\rho}\bigl[ \omega_m^{(\nu)} + \mu_{l+1,i} \bigr] }{ \bigl(
|\mu_{l+1,r}| + \bigl(0.5\omega_0^2\rho^2c\bigr) \bigr)^2 + \bigl(
\omega_m^{(\nu)} + \mu_{l+1,i} \bigr)^2}   \label{sec1:eq16}
\end{equation}

with $\omega_0 = 2\pi/T_0$ being the operating frequency of the synchronized
ensemble, $\mu_s = \mu_{s,r} + j \mu_{s,i} \in \mathbb{C}$, is the $s$th
characteristic Floquet exponent (real/imaginary parts) and
$\omega^{(\nu)}_m = \omega - \nu\omega_0$ is the $\nu$th harmonic
offset frequency. The real parameter, $c \in \mathbb{R}_{+}$, is known as the 
\emph{phase-diffusion} constant\cite{Demir2000,Traversa2011,Djurhuus2009} 
(see also \cref{sec1:sub1:sub2}). All summations in \cref{sec1:eq16} without bounds operate in the interval $(-\infty,\infty)$.  
The operators $a^{(\nu)},b^{(\nu)},\Upsilon^{(\nu)}_{l\rho},\Delta^{(\nu)}_{l\rho}$ introduced 
in \cref{sec1:eq16} are defined through\footnote{
Let $M^{(\nu)} \in \mathbb{C}^{n\times n}$ be some square, complex 
matrix (within the COSC-PMM methodology) with superscript $\nu$ referring to the PSS harmonic index discussed in connection 
with \cref{sec1:eq16}. Then let, ${}_{q}\overline{M^{(\nu)}} =  
[M^{(\nu)}]_{q,q}/\Vert X^{[i]}_{s,\nu}\Vert^2$, 
where $[Q]_{i,j}$ denotes to the 
element at row $i$ and column $j$ of matrix $Q$ and $X^{[q]}_{s,\nu}$ refers 
to the $q$th element of the PSS $\nu$th harmonic vector. Hence, 
${}_{q}\overline{M^{(\nu)}}$ is the $q$th diagonal 
matrix element, normalized by the $\nu$th harmonic power, of the $q$th PSS vector 
component. \label{sec1:foot2}} $a^{(\nu)}+jb^{(\nu)} = {}_{q}\overline{\Omega^{(\nu)}}$ and 
$\Upsilon^{(\nu)}_{l\rho} + j \Delta^{(\nu)}_{l\rho} = 
 {}_{q}\overline{\Theta_{l\rho}^{(\nu)}}$, where index $q$ refers to the index of the 
 \emph{observation-node} \emph{i.e.} the node where PNOISE is measured (see \cref{sec1:foot2}).   
The two complex tensor operators $\Omega$ and $\Theta_{l\rho} \in \mathbb{C}^{n\times n}$, 
used in the above definitions, have the form 
(see \cite[appendix D]{Djurhuus2022} for details)

\begin{align}
\Omega^{(\nu)} &=  \sum_{m=2}^k \sum_{p} \frac{
  U_{1,\nu}\Lambda_{1,0}^{\top}\Lambda_{m,\nu-p}^*U_{m,p}^{\dagger}}{j\omega_0(p-\nu)
  - \mu_m^*} \label{sec1:eq18a}\\
  \Theta_{l\rho}^{(\nu)} &= \frac{
  U_{1,\rho}\Lambda_{1,0}^{\top}\Lambda_{l+1,\rho-\nu}^*U_{l+1,\nu}^{\dagger}}{j\omega_0(\nu-\rho)
    - \mu_{l+1}^* - \mu_1} +  \sum_{i = 2}^k\sum_p\frac{
    U_{i,p}\Lambda_{i,\rho-p}^{\top}\Lambda_{l+1,\rho-\nu}^*U_{l+1,\nu}^{\dagger}}
    {j\omega_0(\nu-p)
    - \mu_{l+1}^* - \mu_i}\label{sec1:eq18b}
\end{align}

where $U_{i,j} \in \mathbb{C}^n$ is the $j$th harmonic of the $i$th 
Floquet vector $u_i(t) : \mathbb{R} \to \mathbb{C}^n$ 
and $\Lambda_{i,j} \in \mathbb{C}^n$ is the $j$th harmonic of the $i$th 
Floquet lambda-vector, $\lambda_i(t) : \mathbb{R} \to \mathbb{C}^n$, 
where $\lambda_i(t) = v_i^{\top}(t)B(x_s(t))$, with $v_i(t) : \mathbb{R} \to \mathbb{C}^n$ 
being the $i$th Floquet phase-mode dual-vector (see \cref{sec1:sub1}). Finally, $B(x_s(t)) : \mathbb{R} \to \mathbb{R}^{n\times p}$ is the noise modulation 
matrix, see \cite{Kaertner1990,Demir2000,Traversa2011,Djurhuus2022} and \cref{sec1:eq1}, 
which describes how $p$ unit-power, zero-mean white-noise sources are injected into the $n$-dimensional 
state-space, the power of these sources and the manner in which the sources are modulated by 
the PSS.

\subsection{Expanding the COSC-PMM Framework : Adding Amplitude and Correlation Contributions.}

\label{sec1:sub3}

Using the definition of residual phase and amplitude response in \cref{sec1:eq14} 
the complete LR state-vector is written   

\begin{equation}
x_{\text{LR}}(t)|_{\mathcal{M}} = x_s(\iota(t)) + \Phi_{\delta}(\iota(t))  + A_{\delta}(\iota(t)) 
\label{sec1:eq19}
\end{equation}

From \cref{sec1:eq19}, the system correlation matrix, $C_{s}(\tau) = 
\lim_{t\to \infty}\bigl< x_{\text{LR}}(t)x_{\text{LR}}^{\dagger}(t+\tau)\bigr>$, 
is decomposed as follows

\begin{equation}
C_{s}(\tau) = C_{\Phi\Phi}(\tau) + C_{\Phi A}(\tau) +C_{AA}(\tau) \label{sec1:eq20}
\end{equation}

where,  $C_{\Phi\Phi}, C_{\Phi A}, C_{AA} : \mathbb{R} \to \mathbb{R}^{n\times n}$, 
represent the COSC-PMM asymptotic phase, cross-correlation and amplitude correlation matrices for 
the synchronized $k$-ensemble, respectively, with

\begin{align}
C_{\Phi\Phi}(\tau) &= \lim_{t\to\infty}\bigl< \bigl[x_s(\iota(t)) + \Phi_{\delta}(\iota(t)) \bigr]
\bigl[x_s(\iota(t+\tau)) 
+ \Phi_{\delta}(\iota(t+\tau)) \bigr]^{\dagger} \bigr> \label{sec1:eq21a}\\
C_{\Phi A}(\tau) &= \lim_{t\to\infty}\bigl<x_s(\iota(t))A_{\delta}^{\dagger}(\iota(t+\tau))\bigr> + 
\lim_{t\to\infty}\bigl<\Phi_{\delta}(\iota(t))A_{\delta}^{\dagger}(\iota(t+\tau))\bigr> + 
\bigl<\cdots\bigr>^{\dagger} \label{sec1:eq21b} \\
C_{AA}(\tau) &= \lim_{t\to\infty}\bigl< A_{\delta}(\iota(t))A_{\delta}^{\dagger}(\iota(t+\tau)) \bigr> \label{sec1:eq21c}
\end{align}

Fourier transforming the $q$th diagonal component, of the $\nu$th harmonic 
envelope, of \cref{sec1:eq21a} produces an expression for 
the $k$-ensemble phase-noise spectrum, around the $\nu$ harmonic and for the $q$th measurement node, 
which was done in \cite{Djurhuus2022} and with this result repeated in \crefrange{sec1:eq16}{sec1:eq18b}. 
Similarly, components of the correlation matrices in \cref{sec1:eq21b,sec1:eq21c} can be 
transformed to produce the cross-correlation 
and amplitude spectra, $\mathcal{R}^{(\nu)},\mathcal{A}^{(\nu)} : 
\mathbb{R} \to \mathbb{R}$.

\subsubsection{The COSC-PMM Amplitude Noise Spectrum.}
\label{sec1:sub3:sub1}

Inspecting \cref{sec1:eq21c,sec1:eq14}, it follows that calculating 
the asymptotic amplitude correlation matrix, $C_{AA}(\tau) : \mathbb{R} \to \mathbb{C}^{n \times n}$, 
involves evaluating terms of the type $\bigl< \psi_{i}(t)\psi_{j}(t+\tau)\bigr>$, 
where, $\psi_i: \mathbb{R} \to \mathbb{C}^n$, is the $i$th amplitude Floquet mode as defined above 
in connection with \cref{sec1:eq14} (see also \cref{sec1:sub1:sub1,sec1:foot4}). The techniques used for 
calculating this type of ensemble average was thoroughly covered in 
\cite{Djurhuus2022}. Following the procedure laid out in appendix B of \cite{Djurhuus2022} an 
expression for the amplitude spectrum $\mathcal{A}^{(\nu)} : 
\mathbb{R} \to \mathbb{R}$, can be derived  

\begin{equation}
\mathcal{A}^{(\nu)}(\omega_m^{(\nu)}) =  
\sum_{\rho} \sum_{l=1}^{n{-}k} \frac{ W^{(\nu)}_{l\rho} \bigl[
2|\mu_{l+k,r}| + (\omega_0^2\rho^2c) \bigr] +
2T^{(\nu)}_{l\rho}\bigl[ \omega_m^{(\nu)} + \mu_{l+k,i} \bigr] }{ \bigl(
|\mu_{l+k,r}| + \bigl(0.5\omega_0^2\rho^2c\bigr) \bigr)^2 + \bigl(
\omega_m^{(\nu)} + \mu_{l+k,i} \bigr)^2} \label{sec1:eq22}
\end{equation}

where 
$W_{l\rho}^{(\nu)}+jT_{l\rho}^{(\nu)} =  {}_{q}\overline{\Pi_{l\rho}^{(\nu)}}$ (see \cref{sec1:foot2}) 
and

\begin{equation}
\Pi_{l\rho}^{(\nu)} = 
   \sum_{i = k+1}^n  \sum_p\frac{
    U_{i,p}\Lambda_{i,\rho-p}^{\top}\Lambda_{l+k,\rho-\nu}^*U_{l+k,\nu}^{\dagger}}{j\omega_0(\nu-p)
    - \mu_{l+k}^* - \mu_i} \label{sec1:eq23}
\end{equation}

\subsubsection{The COSC-PMM Cross-Correlation Noise Spectrum}
\label{sec1:sub3:sub2}

The amplitude-phase cross-correlation matrix is written in \cref{sec1:eq21b} 
with the various operators defined in \cref{sec1:eq14,sec1:eq19}. It involves 
calculating terms of the type $\bigl<x_s(t + \alpha(t))\psi_i^{\dagger}(t+\tau)\bigr>$
and $\bigl<\psi_{i}(t)\phi_j^{\dagger}(t+\tau)\bigr>$ 
where $\phi_i,\psi_i$ represent the $i$th residual phase and 
amplitude Floquet modes, respectively. 
Again, following the procedure laid out in appendix B of \cite{Djurhuus2022}, 
which explains how to calculate both of these two types ensemble averages, 
an expression for the cross-correlation spectrum $\mathcal{R}^{(\nu)} : 
\mathbb{R} \to \mathbb{R}$, can be derived  

\begin{equation}
\mathcal{R}^{(\nu)}(\omega_m^{(\nu)}) =  
-\frac{ t^{(\nu)}(\omega_0^2\nu^2c)
+ 2u^{(\nu)} \omega_m }{ (0.5\omega_0^2\nu^2c)^2 + (\omega_m^{(\nu)})^2} + 
\sum_{\rho} \sum_{l=1}^{n-1} \frac{ E^{(\nu)}_{l\rho} \bigl[
2|\mu_{l+1,r}| + (\omega_0^2\rho^2c) \bigr] +
2Z^{(\nu)}_{l\rho}\bigl[ \omega_m^{(\nu)} + \mu_{l+1,i} \bigr] }{ \bigl(
|\mu_{l+1,r}| + \bigl(0.5\omega_0^2\rho^2c\bigr) \bigr)^2 + \bigl(
\omega_m^{(\nu)} + \mu_{l+1,i} \bigr)^2}   \label{sec1:eq24}
\end{equation}

where , 
$t^{(\nu)}+ju^{(\nu)} = {}_{q}\overline{\Psi^{(\nu)}}$,
$E^{(\nu)}_{l\rho} + j Z^{(\nu)}_{l\rho} =  {}_{q}\overline{\Xi^{(\nu)}_{l\rho}}$ (see \cref{sec1:foot2}) and

\begin{align}
  \Psi^{(\nu)} &=  \sum_{m=k+1}^n \sum_{p} \frac{
  U_{1,\nu}\Lambda_{1,0}^{\top}\Lambda_{m,\nu-p}^*U_{m,p}^{\dagger}}{j\omega_0(p-\nu)
  - \mu_m^*} \label{sec1:eq24a}\\
  \Xi_{l\rho}^{(\nu)} &=  \begin{cases}
    \sum_{i = k+1}^n  \sum_p\frac{
    U_{i,p}\Lambda_{i,\rho-p}^{\top}\Lambda_{l+1,\rho-\nu}^*U_{l+1,\nu}^{\dagger}}{j\omega_0(\nu-p)
    - \mu_{l+1}^* - \mu_i} \quad &\text{for $l \leq k-1$} \\
    \frac{
  U_{1,\rho}\Lambda_{1,0}^{\top}\Lambda_{l+1,\rho-\nu}^*U_{l+1,\nu}^{\dagger}}{j\omega_0(\nu-\rho)
    - \mu_{l+1}^* - \mu_1} +  \sum_{i = 2}^k  \sum_p\frac{
    U_{i,p}\Lambda_{i,\rho-p}^{\top}\Lambda_{l+1,\rho-\nu}^*U_{l+1,\nu}^{\dagger}}{j\omega_0(\nu-p)
    - \mu_{l+1}^* - \mu_i} \quad &\text{for $l > k-1$} \label{sec1:eq24b}
  \end{cases}
\end{align}

\subsubsection{The Reduced Single Oscillator Expressions. }
\label{sec1:sub3:sub3}

The equivalent spectral densities for a single/free-running oscillator 
are found from the above expressions by substituting $k=1$. The phase, amplitude and cross-correlation 
noise spectra defined above in \cref{sec1:eq16,sec1:eq22,sec1:eq24} are then reduced to the following 
single/free-running oscillator expressions

\begin{align}
\mathfrak{L}^{(\nu)}(\omega_m^{(\nu)}) &= 
\frac{ \omega_0^2\nu^2c}{ (0.5\omega_0^2\nu^2c)^2 + (\omega_m^{(\nu)})^2}  \label{sec1:eq25a}\\
\mathcal{A}^{(\nu)}(\omega_m^{(\nu)}) &=  
\sum_{\rho} \sum_{l=1}^{n-1} \frac{ W^{(\nu)}_{l\rho} \bigl[
2|\mu_{l+1,r}| + (\omega_0^2\rho^2c) \bigr] +
2T^{(\nu)}_{l\rho}\bigl[ \omega_m^{(\nu)} + \mu_{l+1,i} \bigr] }{ \bigl(
|\mu_{l+1,r}| + \bigl(0.5\omega_0^2\rho^2c\bigr) \bigr)^2 + \bigl(
\omega_m^{(\nu)} + \mu_{l+1,i} \bigr)^2} \label{sec1:eq25b}\\
\mathcal{R}^{(\nu)}(\omega_m^{(\nu)}) &=  
-\frac{ t^{(\nu)}(\omega_0^2\nu^2c)
+ 2u^{(\nu)} \omega_m }{ (0.5\omega_0^2\nu^2c)^2 + (\omega_m^{(\nu)})^2} + 
\sum_{\rho} \sum_{l=1}^{n-1} \frac{ E^{(\nu)}_{l\rho} \bigl[
2|\mu_{l+1,r}| + (\omega_0^2\rho^2c) \bigr] +
2Z^{(\nu)}_{l\rho}\bigl[ \omega_m^{(\nu)} + \mu_{l+1,i} \bigr] }{ \bigl(
|\mu_{l+1,r}| + \bigl(0.5\omega_0^2\rho^2c\bigr) \bigr)^2 + \bigl(
\omega_m^{(\nu)} + \mu_{l+1,i} \bigr)^2} \label{sec1:eq25c}
\end{align}

where the various operators are defined above in \cref{sec1:sub3:sub1,sec1:sub3:sub2} 
with $k=1$. The results in \crefrange{sec1:eq25a}{sec1:eq25c} agree 
with the prior published results in \cite{Kaertner1990,Demir2000,Traversa2011}. Specifically, 
the reduced (single oscillator) PNOISE expression in \cref{sec1:eq25a} represents 
simple Lorentzian functional characteristic of the offset frequency\footnote{ 
Consider the spectrum around the first harmonic, $\nu = 1$, with 
simplified notation $\mathfrak{L}(\omega_m) \equiv \mathfrak{L}^{(1)}(\omega_m^{(1)})$.
From \cref{sec1:eq25a}, $\mathfrak{L}(\omega_m) \approx  
(0.5\omega_0^2c)/ \omega_m^2$ for 
$\omega_m \gg 0.5\omega_0^2c$ which holds for most offsets 
of interest as $c \ll 1$. Thus, for most offset of interest, 
$10\log_{10}(\mathfrak{L}(\omega_m)) \approx 10\log_{10}(0.5\omega_0^2c) -20\log_{10}(\omega_m)$, 
\emph{i.e.} a straight line on a semi-log plot, which drops 20dB/decade. \label{sec1:foot5}}, 
a well-established result\cite{Kaertner1990,Demir2000}. 
This outcome, once again, underscore the important fact, that the novel 
COSC-PMM framework not only extends the earlier published single oscillator 
PMM scheme but also \underline{entirely replaces it}. A notable innovation, resulting from 
the COSC-PMM framework, has been the complete recalibration of how we 
approach time-domain macro-modelling of oscillator noise-response. It is now understood that 
the established PMM framework (see \cite{Kaertner1990,Demir2000,Traversa2011}), 
while representing an extremely impressive theoretical achievement, 
is indeed, without hyperbole, incomplete. The PMM represents a scope limited description 
which only covers a very minute subset of possible relevant circuits; namely the subset of 
free-running oscillators/clocks in the set of all oscillating circuits. 
The novel COSC-PMM methodology, on the other hand, is complete, and it 
replaces the old PMM description by extending the model domain to 
include the \underline{entire set} of all possible oscillating autonomous networks.

\subsection{Implementing the \texttt{QUCS-COPEN} PNOISE Module.}
\label{sec1:sub4}

\begin{algorithm}
\small

\caption{\texttt{QUCS-COPEN} PNOISE Module Kernel.}\label{sec1:alg1}
\textbf{Parameters :} $\texttt{Start}, \texttt{Stop}, \texttt{Type},\texttt{N}$ 
\Comment{freq. swp range, type [\texttt{lin/log}] \& \# points(/dec)} \\
\textbf{Parameters :} $\texttt{Nf},\texttt{NoiseNodes}$,    
\Comment{\texttt{Nf} : \# harmonics, \texttt{NoiseNodes} : output node-names. } \\
\textbf{Parameters :} k, $\texttt{PssSim}$  
\Comment{k : \# osc. units, \texttt{PssSim} : PSS ref. obj. name.} \\
\textbf{Input :} $x_{\text{\tiny PSS}}(t)$  \Comment{calculated PSS solution.}\\
\textbf{Output Datasets :} pnoise (Xpn), xcorr (Xcr), anoise (Xan)
\Comment{DS file ext. with X=[V/I], Volt./Current.}
\begin{algorithmic}[1]
\Require $\texttt{Start}>0,\texttt{Stop} > \texttt{Start}, \texttt{N(Type=lin|log)} \geq 10|3, \texttt{Nf} \geq 16$
\Ensure $\texttt{isNodeArray}(\texttt{noiseNodes})=\texttt{true}$ \Comment{check : all input nodes exist in circ.}
\Ensure $\texttt{isNum}(x_{\text{\tiny PSS}}) \AND \texttt{isPeriodic}(x_{\text{\tiny PSS}}) = \texttt{true}$ \Comment{check : $x_{\text{PSS}}$ is valid PSS solution.}
\Ensure $\texttt{moduleExists}(\texttt{PssSim})=\texttt{true}$ \Comment{check : reference PSS module exist in schematic.}
\algrule[0.8pt]
\State \texttt{SubProc.\,I} : Calculate Floquet Decomposition. \label{sec1:alg1:proc1}
\algrule[0.8pt]
\State $\Phi_{\text{\tiny MM}} \gets \texttt{CalcMonodromy}(T_0;x_{\text{PSS}},\texttt{{\small ADJOINT}=false})$
\Comment{calc. standard monodromy from \cref{sec1:eq4}.}
\State $\Psi_{\text{\tiny MM}} \gets \texttt{CalcMonodromy}(T_0;x_{\text{PSS}},\texttt{{\small ADJOINT}=true})$
\Comment{calc. adjoint monodromy from \cref{sec1:eq8}.}
\State $\{u_{i0}, \mu_i\}_{i=1}^n \gets \texttt{CalcEigenSet}( \Phi_{\text{\tiny MM}})$
\State $\{v_{i0}, \mu_i\}_{i=1}^n \gets \texttt{CalcEigenSet}( \Psi_{\text{\tiny MM}})$
\State $B(t) \gets \texttt{NoiseMatrixSeries}(x_{\text{PSS}})$
\State $L \gets \texttt{NumberRelevantModes}(\{\mu_i\},T_0)$ \Comment{ count \# relevant Floquet modes.}
\label{sec1:alg1:relevantModes}
\For { $j = 1 \to L$}
\State $u_{j}(t) \gets \texttt{LrIntegrate}(u_{j0},\texttt{{\small ADJOINT}=false})$ \Comment{integrate \cref{sec1:eq2} {\bf fwd} in time with init. cond. $u_{j0}$.} 
\State $v_{j}(t) \gets \texttt{LrIntegrate}(v_{j0},\texttt{{\small ADJOINT}=true})$ \Comment{integrate \cref{sec1:eq6} {\bf bwd} in time with init. cond. $v_{j0}$.}
\State $\lambda_j(t) \gets \texttt{generateLambda}(B(t),v_j(t))$ 
\Comment{calculate : $\lambda_j(t) = v_j^{\top}(t)B(t)$} \label{sec1:alg1:lambda}
\State $U_j \gets \texttt{VectorFFT}(u_{j}(t),\texttt{Nf})$
\State $\Lambda_j \gets \texttt{VectorFFT}(\lambda_{j}(t),\texttt{Nf})$
\EndFor

\algrule[0.8pt]
\State \texttt{SubProc.\,II} : Calculate Noise Operators. \label{sec1:alg1:proc2}
\algrule[0.8pt]

\State $c \gets \texttt{CalcPhaseDiff}(\Lambda_1)$ \Comment{calc. phase diff. const., see \cref{sec1:sub1:sub2}.}
\State $\texttt{params} \gets \texttt{struct}(k,\nu,c,T_0,\texttt{Nf},L,\{U_i,\mu_i,\Lambda_i\})$

\For{ ndName in noiseNodes }

\State $q \gets \texttt{NodeName2Index}( \text{ndName} )$ \Comment{$q$ : state vec. index corresp. to ndName.}

\State ${}_{q}\Omega^{(\nu)} \gets \texttt{CalcOmegaMatrix}(\texttt{q,params})$ 
\Comment{eval. $q$th diagonal comp. of \cref{sec1:eq18a}}
\State ${}_{q}\Upsilon^{(\nu)} \gets \texttt{CalcUpsilonMatrix}(\texttt{q,params})$ 
\Comment{eval. $q$th diagonal comp. of \cref{sec1:eq24a}}

\For{ $\rho = -\texttt{Nf} \to \texttt{Nf}$}
\For { $l = 1 \to k-1$ }

\State ${}_{q}\Theta_{l\rho}^{(\nu)} \gets \texttt{CalcThetaTensor}(l,\rho,q,\texttt{params})$ 
\Comment{ \cref{sec1:eq18b} @ index $(l,\rho)$, ($q$th diag. comp.).}
\EndFor

\For { $l = 1 \to L-k$ }
\State ${}_{q}\Pi_{l\rho}^{(\nu)} \gets \texttt{CalcPiTensor}(l,\rho,q,\texttt{params})$  
\Comment{\cref{sec1:eq23} @ index $(l,\rho)$, ($q$th diag. comp.).}
\EndFor

\For { $l = 1 \to L$ }
\State ${}_{q}\Xi_{l\rho}^{(\nu)} \gets \texttt{CalcXiTensor}(l,\rho,q,\texttt{params})$ 
\Comment{\cref{sec1:eq24b} @ index $(l,\rho)$, ($q$th diag. comp.).}
\EndFor

\EndFor
\EndFor

\algrule[0.8pt]
\State \texttt{SubProc.\,III} : Calculate Noise Spectra \& Save Datasets. \label{sec1:alg1:proc3}
\algrule[0.8pt]

\State $\text{Fsweep} \gets \texttt{GenerateSweep}(\texttt{Start},\texttt{Stop},\texttt{Type},\texttt{N})$

\State $\texttt{params} \gets \texttt{struct}(k,\nu,T_0,\texttt{Nf})$
\For{ ndName in noiseNodes }
\State $q \gets \texttt{NodeName2Index}( \text{ndName} )$
\State $i \gets 0$
\For { fm in Fsweep } 
\State $\text{pnoise}[i] \gets \texttt{calcPnoise}(q,\text{fm},\texttt{params},\Omega^{(\nu)}, \Theta_{l\rho}^{(\nu)})$ 
\Comment{eval. phase noise spectrum, see \cref{sec1:eq16}}
\State $\text{anoise}[i] \gets \texttt{calcAnoise}(q,\text{fm},\texttt{params},\Pi_{l\rho}^{(\nu)})$ 
\Comment{eval. amp. noise spectrum, see \cref{sec1:eq22}}
\State $\text{xnoise}[i] \gets \texttt{calcXnoise}(q,\text{fm},\texttt{params},\Upsilon^{(\nu)}, \Xi_{l\rho}^{(\nu)})$ 
\Comment{eval. xcorr noise spectrum, see \cref{sec1:eq24}}
\State $i \gets i+1$
\EndFor
\State $\texttt{SaveDatasets}(\text{ndName},\text{Fsweep},\text{pnoise},\text{anoise},\text{xnoise})$ 
\Comment{Store ndName datasets for print.} 

\EndFor

\end{algorithmic}
\end{algorithm}

The \texttt{QUCS-COPEN} PNOISE module kernel algorithm is outlined in \cref{sec1:alg1}.
The module code is encapsulated within the new \texttt{pnsolver} 
C++ class which is part of the modified \texttt{qucsator} engine codebase (see \cite{Djurhuus2025} and \cref{sec0:foot1}). 
The \texttt{pnsolver} C++ class inherits directly from the 
established \texttt{nasolver} class, which represent the base-class interface 
to all \texttt{QUCS} analysis classes. This new PNOISE module introduces a novel, previously absent, 
large-signal noise analysis utility to the \texttt{qucsator/QUCS} API.
\par
Each instance of the \texttt{pnsolver} analysis class is given 
a reference to a single corresponding \texttt{psssolver} instance/object which is 
assumed to produce the large-signal PSS solution required to calculate the noise-response 
(see \cref{sec1:sub1,sec1:sub2}). The implementation of the 
\texttt{psssolver} object class was discussed in \cite{Djurhuus2025}. 
The \texttt{psssolver} 
reference object is identified through the string 
input parameter, $\texttt{PssSim}$, and the program fails to execute if no instance 
with this name exists in the schematic. The main routine, 
outlined in \cref{sec1:alg1}, can be subdivided into three sub-procedures, referred to 
as \texttt{SubProc.\,I-III}. 

\subsubsection{\texttt{SubProc.\,I} : the Floquet Decomposition.}
\label{sec1:sub4:sub1}

Starting on \cref{sec1:alg1:proc1} in \cref{sec1:alg1}, this portion 
of the code is concerned with calculating the spectral Floquet operators 
$\{U_{i,j},\Lambda_{i.j}\}$, 
introduced in \cref{sec1:sub2} and which are needed in-order 
to calculate the various noise operators in 
\cref{sec1:eq16,sec1:eq18a,sec1:eq18b,sec1:eq22,sec1:eq23,sec1:eq24,sec1:eq24a,sec1:eq24b}. 
In-order to achieve this goal the corresponding 
time-domain Floquet objects (see \cref{sec1:sub1}), $u_i(t)$, and $\lambda_i(t) = v_i^{\top}(t)B(x_s(t))$ are first derived. 
\par
First, the (adjoint) monodromy matrix (MM) is evaluated through 
a call to \texttt{CalcMonodromy}, with the switch 
\texttt{\small ADJOINT} controlling whether the standard or adjoint MM is derived 
using the formulae in \cref{sec1:eq4} or \cref{sec1:eq8}, respectively\footnote{
Note, that when integrating the (adjoint) LR equations in \cref{sec1:eq2,sec1:eq6}, 
for the purpose of deriving  \cref{sec1:eq4,sec1:eq8}, the LR equations should be 
integrated using the same Linear-Multi-Step (LMS) 
setup as was used for calculating the PSS 
(see discussion in \cite{Djurhuus2025}). Also note that the adjoint equations 
must be integrated backwards in time \cite{Demir2000,Demir1998}. \label{sec1:foot3}}. The (dual) 
Floquet vectors, $\{u_{i0},v_{i0}\}$, and 
Floquet exponents, $\{\mu_i\}$, are then 
derived from the eigen-decomposition (note that $\mu_i = \log(\lambda_i)/T_0$, 
see \cref{sec1:sub1}) of this (adjoint) MM using a standard QR Shur-shift 
algorithm routine, referred to in \cref{sec1:alg1} as \texttt{CalcEigenSet}. 
\par
Some of the calculated 
Floquet modes are dampened to an extent where they do not 
contribute noticeably to the spectral output 
at any offset above the noise-floor including 
the modes spanning the null-space of the (potentially) singular 
$C$ Jacobian matrix (see \cref{sec1:eq1,sec1:eq2}), which corresponds 
to modes with negative infinite Floquet 
exponents (real part), $\lambda_i = 0 \Rightarrow \Re\{\mu_i\} = -\infty$. Since these types of 
modes do not contribute to the visible part of the calculated spectrum (above the noise floor) 
is it only natural to neglect these contributions. Doing so will 
shrink the size of the effective Floquet set and thus significantly improve 
the computational efficiency of the algorithm. From \cite{Maffezzoni2013}, the 
so-called, \emph{relevant modes}, satisfy $|\Re\{\mu_i\}| \ll \omega_0$ and for 
our purposes this translates into the rule $\Re\{\mu_i\} \leq 10\omega_0$. On \cref{sec1:alg1:relevantModes} 
of \cref{sec1:alg1} the call to \texttt{NumberRelevantModes} counts these modes, 
according to the mentioned rule, thus allowing us to shrink the effective 
Floquet set to from $n$ to $L \leq n$ modes. Next, the noise-matrix, $B(x_s(t))$, defined in \cref{sec1:eq1}, 
is then evaluated by probing the circuit noise source contributions along the PSS, $x_s(t)$. 
This is done in \cref{sec1:alg1} with a call to \texttt{NoiseMatrixSeries}. 
\par
The time-series of the $j$th (dual) Floquet vector $u_j(t),v_j(t)$ is derived 
by integrating the (adjoint) LR equations in 
\cref{sec1:eq2} and \cref{sec1:eq6}, respectively with initial conditions, 
$\{u_{j0},v_{j0}\}$, (backwards) forwards in time. These calculations are done with a 
call to \texttt{LRintegrate} with the switch 
\texttt{\small ADJOINT} controlling whether the standard LR equations in \cref{sec1:eq2} 
or the adjoint version in \cref{sec1:eq5} are invoked. Finally, the lambda-operators can be formed (\texttt{generateLambda} on \cref{sec1:alg1:lambda}) and 
the corresponding spectral operators, mentioned above, are evaluated by a call to a standard FFT 
library routine.

\subsubsection{\texttt{SubProc.\,II + III} : Calculate Noise Operators \& Output Spectra.}
\label{sec1:sub4:sub2}
The second part of \cref{sec1:alg1} 
starts on \cref{sec1:alg1:proc2}. It concerns the calculation 
of the noise operators defined in \cref{sec1:sub2,sec1:sub3}. 
Expressions such as \emph{e.g.} \cref{sec1:eq18a,sec1:eq18b,sec1:eq23} 
\emph{etc.}, all involve infinite sums, represented in the equations as 
summations without limits. An example hereof could be the summation, $\sum_{p}$, in \cref{sec1:eq24a}, 
with $p \in ]-\infty;\infty[$. The index variables, referenced in these infinite sums, 
all label harmonics of Fourier 
transformed operators. Hence, as written in \emph{e.g.} \cref{sec1:eq18a,sec1:eq18b,sec1:eq23}, 
these expressions all assume infinite harmonic contributions. Obviously, for the algorithm 
considered here, the range of these sums must be capped. Thus, in \cref{sec1:alg1}, 
all harmonic sums are restricted to the domain, $[-\texttt{Nf},\texttt{Nf}]$, where \texttt{Nf} is an input parameter 
(see parameter section of \cref{sec1:alg1}).  
\par
The main procedure involves several nested for loops with the outermost loop iterating 
over the set of observation/measurement nodes provided by the input parameter 
string-list $\texttt{NoiseNodes}$. From the discussion in \cref{sec1:sub2}, and in 
\cite{Djurhuus2022}, the aim of the algorithm should be to calculate the auto-correlation spectrum 
of a given measurement node. Hence, for a measurement index, $q$, corresponding 
to a circuit node name in $\texttt{NoiseNodes}$, only the 
$q$th diagonal element of the noise-correlation spectrum needs to be calculated. 
This, in-turn, implies that only $q$th diagonal element of the various noise operators, 
discussed here and in \cref{sec1:sub2,sec1:sub2}, must be calculated. For 
the matrices $\Omega^{(\nu)}$ and $\Psi^{(\nu)}$, defined in \cref{sec1:eq18a,sec1:eq24a}, 
this element is written  ${}_q\Omega^{(\nu)}, {}_q\Psi^{(\nu)}$ with the notation explained 
in \cref{sec1:foot2}. Similarly, the 3 tensors 
$\Theta_{l\rho}^{(\nu)}, \Pi_{l\rho}^{(\nu)}, \Xi_{l\rho}^{(\nu)}$, 
defined in \cref{sec1:eq18b,sec1:eq23,sec1:eq24b} are calculated for indices, $l,\rho$, 
where harmonic index $\rho$ runs from $\texttt{-Nf}$ to $\texttt{Nf}$ 
and the domain of index, $l$, depends on the specific operator being calculated.   
\par
The last portion of the QUCS-COPEN PNOISE algorithm in \cref{sec1:alg1} start 
on \cref{sec1:alg1:proc3} and involves calculating 
the various COSC-PMM spectra defined in \cref{sec1:eq16,sec1:eq22,sec1:eq24}. 
To do so, the procedure employs the various 
noise-operators calculated in \texttt{Proc.\,II}. 
The spectra are calculated for each node in the array \texttt{NoiseNodes} and then stored 
to datasets, which are then printed according to internal \texttt{qucsator} procedures.

\subsection{The GUI Environment \& Output Datasets.}
\label{sec1:sub5}

\begin{figure}[!h]
\begin{center}
\includegraphics[scale=\myFigureScale]{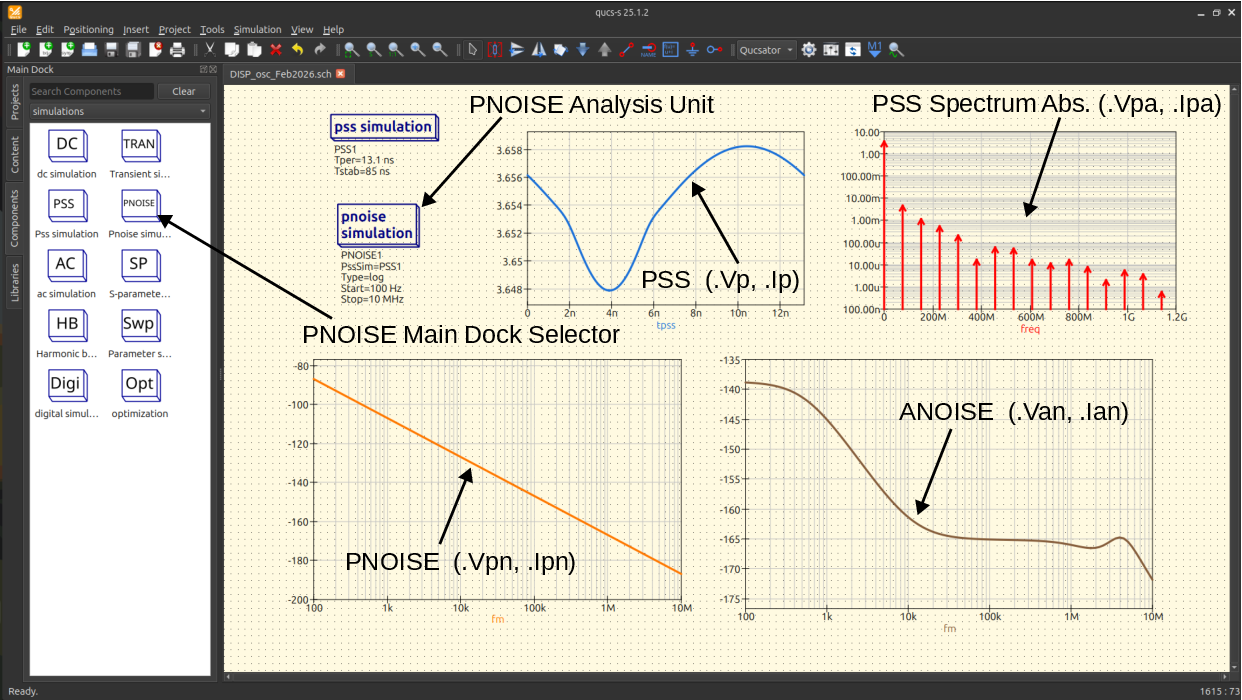}
\end{center}
\caption{The figure shows the \texttt{QUCS-COPEN} 
PSS+PNOISE simulation units integrated into the 
\texttt{QUCS-S} GUI environment. Both the PNOISE and PSS reference modules are chosen 
from the main-dock and then placed on the schematic. 
The PNOISE \& PSS modules are linked through the PNOISE parameter \texttt{PssSim} (see \cref{sec1:alg1}). 
PSS \& PNOISE datasets (unspecified circuit \& output node), with file extensions, 
\textcolor{blue}{blue graph} : PSS time-domain solution, (\texttt{.Vp}, \texttt{.Ip}), 
\textcolor{red}{red graph} : PSS spectrum, absolute value, (\texttt{.Vpa}, \texttt{.Ipa}), 
which is plotted in $\mathrm{dBm}$, \textcolor{orange}{orange graph} : 
Phase noise spectrum, in $\mathrm{dBc}$, (\texttt{.Vpn}, \texttt{.Ipn}) and 
\textcolor{brown}{brown graph} : Amplitude noise spectrum, in $\mathrm{dBc}$ (\texttt{.Van}, \texttt{.Ian}).}
\label{sec1:fig1}
\end{figure}

Various details and issues \emph{w.r.t.} the 
\texttt{QUCS-COPEN} GUI was discussed in \cite{Djurhuus2025}. 
\Cref{sec1:fig1} displays the newly implemented PNOISE tool operating 
inside the \texttt{QUCS-COPEN} GUI application (\texttt{QUCS-S} clone, see \cref{sec0:foot1}) along with the required 
companion PSS module, which is referenced by the PNOISE input 
string parameter \texttt{PssSim} (see \cref{sec1:alg1}). The figure shows the 
result of running a PSS+PNOISE simulation for some 
unspecified oscillator circuit. The PNOISE simulator outputs 
3 datasets (DS) referenced by the corresponding 
file extensions : the phase-noise spectrum $\{\texttt{.Vpn,.Ipn}\}$, 
the amplitude noise spectrum $\{\texttt{.Van,.Ian}\}$, 
and the phase-amplitude cross-correlation noise spectrum (not shown) $\{\texttt{.Vxn,.Ixn}\}$.

\begin{figure}[!h]
\begin{center}
\includegraphics[scale=\myFigureScale]{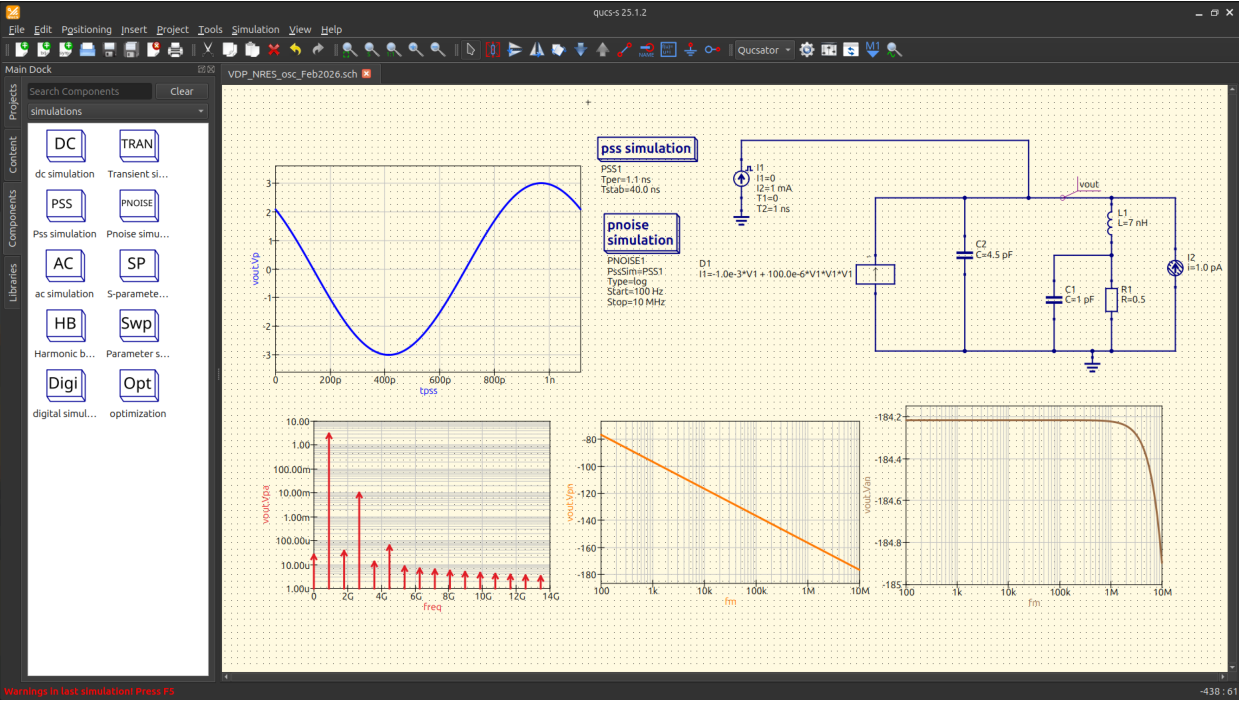}
\end{center}
\caption{The \texttt{QUCS-COPEN} PNOISE calculator module 
applied to a simple Van-der-Pol(VDP)-type oscillator, proposed in 
\cite[Fig. 3]{Djurhuus2022}. This circuit is referred to below as \texttt{OSC.\#1}.
All parameters (including the external noise source) are as in \cite[Fig. 3]{Djurhuus2022} 
(primary osc. parameter set). The datasets introduced in \cref{sec1:fig1} 
are plotted for the voltage node, \texttt{vout}. Refer to \cref{sec1:fig1} 
caption for further details.}
\label{sec2:fig1}
\end{figure}

\section{Example Oscillator Circuits.}
\label{sec2}

Below, the \texttt{QUCS-COPEN} PNOISE module, plus the mandatory companion PSS solver, are applied to three different single/free-running 
oscillator circuits.  Then, in \cref{sec2:sub1}, a coupled oscillator network 
is considered. As far as the authors can tell, the tool demonstrated 
here, performing noise-response analysis of large-signal nonlinear (coupled) oscillating 
circuits, represents the first successful implementation of its kind 
into the \texttt{QUCS} ecosystem. All the circuits considered here, 
with just a few minor modifications (\emph{e.g} addition of external noise sources), 
were also discussed in \cite{Djurhuus2025} in the context of the \texttt{QUCS-COPEN} 
PSS module (steady-state analysis). 
For this reason, only very brief descriptions of these circuits will be given below. For further details, 
please refer to \cite[sec. 3]{Djurhuus2025}.
\par
The first oscillator circuit, shown \cref{sec2:fig1} and known throughout as 
\texttt{OSC.\#1}, is a simple Van-der-Pol equivalent configuration, 
first proposed in \cite{Djurhuus2022}. The PSS module calculates 
a solution with oscillation frequency $f_0 = 1/T_0 = 896.6 \mathrm{MHz}$ and 
the figure plots the relevant PSS \& PNOISE datasets (see \cref{sec1:fig1} caption). 
\Cref{sec2:fig2} displays the same simulation setup 
as in \cref{sec2:fig1}, but this time for a BJT Collpitts oscillator 
first proposed in the paper \cite{Kaertner1990}. Below this circuit is referred 
to as \texttt{OSC.\#2}. 
Compared with the original oscillator design in \cite{Kaertner1990}, the circuit in \cref{sec2:fig2} includes 
a few modifications (see \cite{Djurhuus2025} for details). Finally, \cref{sec2:fig3} shows 
the PNOISE module applied to a MOSFET cross-coupled pair, LC tank oscillator, 
referred to herein as \texttt{OSC.\#3}. The circuit is similar to the MOSFET oscillator proposed in 
\cite[fig. 3]{Maffezzoni2013} including identical device models with a few slight modifications 
(see \cite{Djurhuus2025} for details). 
The PSS simulator finds a solution oscillating 
at, $f_0 = 888.6 \mathrm{MHz}$, and the PSS+PNOISE output datasets, 
introduced in \cref{sec1:fig1}, are plotted.

\subsection{Coupled Oscillator Example Circuit.}
\label{sec2:sub1}

As mentioned in  \cref{sec1:sub2,sec1:sub3:sub3}, the COSC-PMM methodology, introduced in \cite{Djurhuus2022}, both extends and replaces the earlier 
single-oscillator PMM framework, developed in \cite{Kaertner1990,Demir2000,Traversa2011}. 
The replacement claim was directly proven in \cref{sec1:sub3:sub3} and further investigated in the 
free-running oscillator examples discussed above. The new model extends the earlier 
framework by expanding the scope to include \underline{all} oscillating configurations, including 
coupled systems. In this section the \texttt{QUCS-COPEN} PNOISE module is applied to a coupled circuit; 
an injection-locked oscillator (ILO) configuration, 
referred to here \texttt{COSC.\#1}.
\par
The schematic of the \texttt{COSC.\#1} ILO circuit is displayed 
in \cref{sec2:fig4}. It consists of two 
oscillator units, included in the schematic as sub-circuit components, 
of \texttt{OSC.\#1}-type\footnote{Let \texttt{CIR.\#A} represent 
some circuit. Then we say that that \texttt{CIR.\#B} is of \texttt{CIR.\#A}-type if 
\texttt{CIR.\#B} has the same topology (\emph{i.e.} 
same set of components \& connections), but possibly, different 
component values. \label{sec2:foot6}.} (see \cref{sec2:fig1}), coupled 
through a unilateral buffer amplifier. The \emph{primary oscillator}, 
\texttt{PRM-OSC}, injects its output signal, through the unilateral buffer amplifier, 
into the \emph{secondary oscillator}, \texttt{SEC-OSC}. The circuit parameters of the two oscillator 
units are chosen as the primary/secondary values listed in 
\cite[figure 3 caption]{Djurhuus2022}. The coupling buffer is linear with 
trans-conductance $g_m = 50.0 \mathrm{\mu S}$.  The circuit in \cref{sec2:fig4} 
represents a 2-ensemble (see \cref{sec1:sub2}), as it is constructed 
by coupling 2 independent free-running oscillator units which 
implies $k=2$. Here, the COSC-PMM model parameter, $k$, represents the number of oscillator 
units in the coupled ensemble (see \cref{sec1:sub2,sec1:sub3,sec1:alg1}).
\par
\begin{figure}[!h]
\begin{center}
\includegraphics[scale=\myFigureScale]{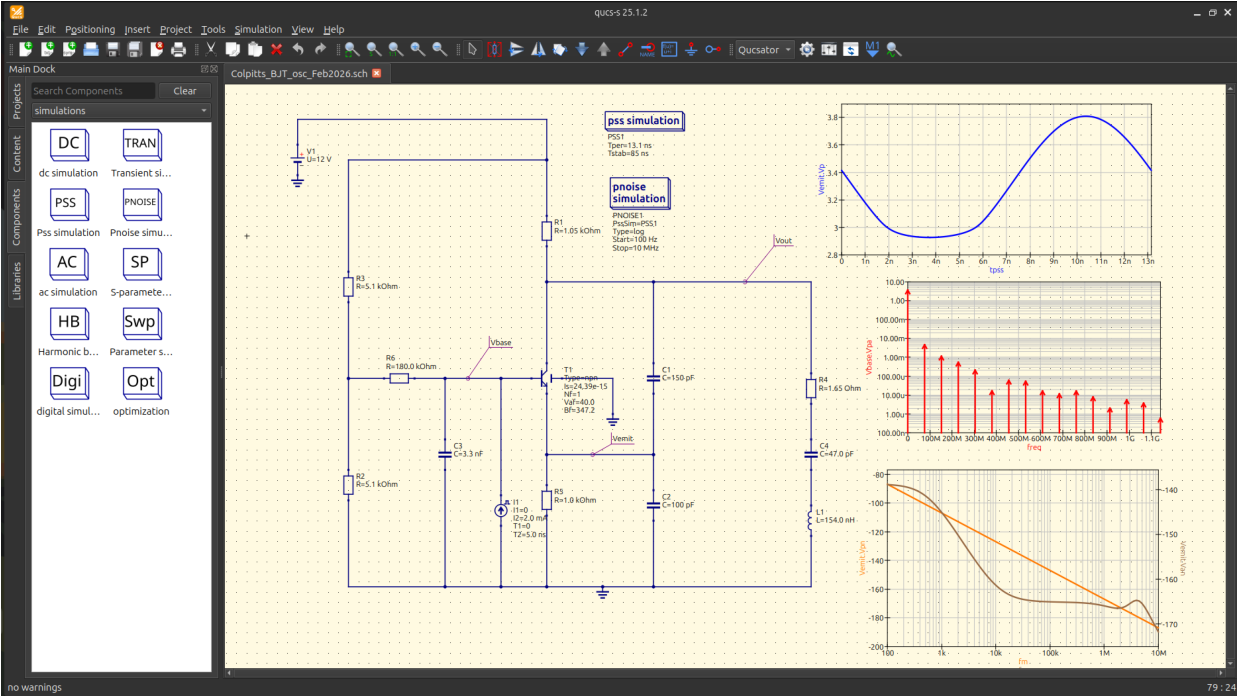}
\end{center}
\caption{ The \texttt{QUCS-COPEN} PNOISE module is applied to a 
BJT Colpitts oscillator taken from \cite{Kaertner1990}, referred to below as \texttt{OSC.\#2}. 
All circuit parameters as in \cite{Kaertner1990}, with a few minor exceptions 
(see \cite{Djurhuus2025} for details). In this simulation the BJT device is noiseless 
except for internal resistor contributions.
The PSS+PNOISE datasets, introduced in \cref{sec1:fig1}, are plotted for the 
BJT emitter node (called \texttt{Vemit} in schematic). See \cref{sec1:fig1} for further details.}
\label{sec2:fig2}
\end{figure}
\begin{figure}[!h]
\begin{center}
\includegraphics[scale=\myFigureScale]{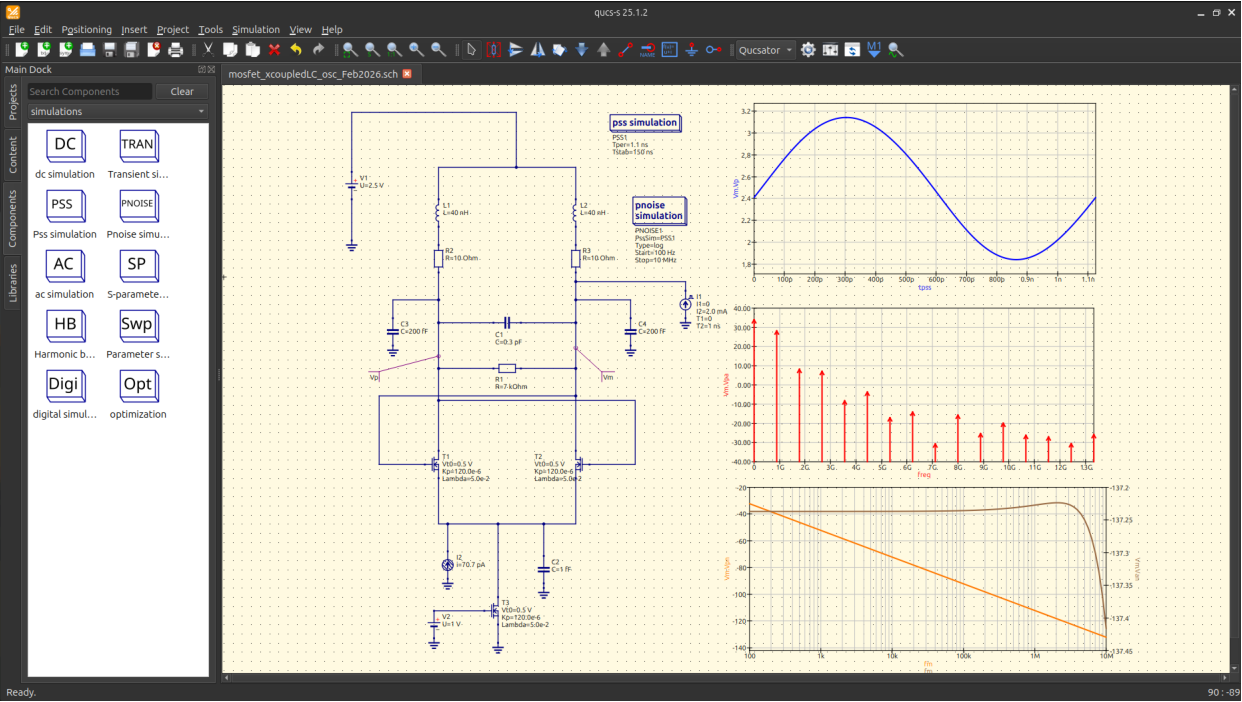}
\end{center}
\caption{The \texttt{QUCS-COPEN} PNOISE simulation tool applied to a 
MOSFET cross-coupled LC-tank oscillator, referred to below as \texttt{OSC.\#3}, similar to the 
oscillator proposed in \cite[fig. 3]{Maffezzoni2013} with minor modifications(see \cite{Djurhuus2025} for details). 
All parameters (including the external noise source) are 
as in \cite[Fig. 3]{Djurhuus2022}. The PSS \& PNOISE datasets, introduced in \cref{sec1:fig1}, 
are plotted for the right MOSFET right drain node \texttt{Vm}.
See also \cref{sec1:fig1} for further details. }
\label{sec2:fig3}
\end{figure}
\begin{figure}[!h]
\begin{center}
\includegraphics[scale=\myFigureScale]{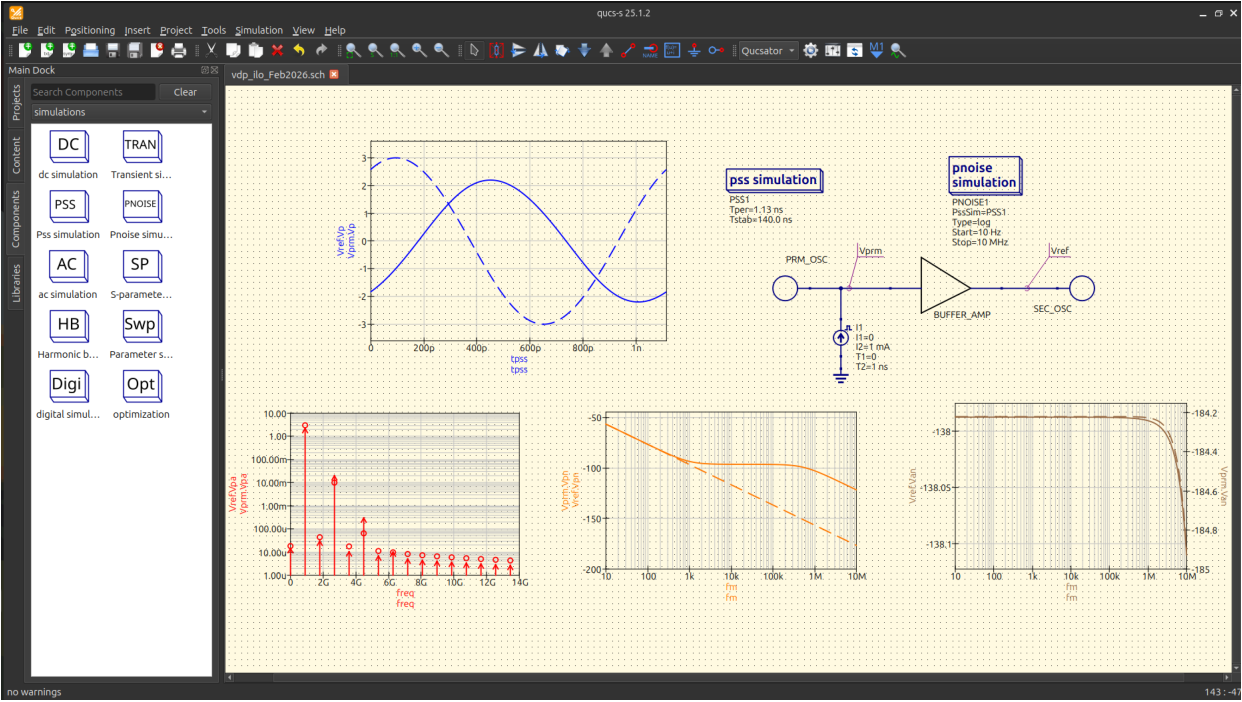}
\end{center}
\caption{ The \texttt{QUCS-COPEN} PNOISE module is applied to an 
injection-locked circuit (ILO) configuration referred to as \texttt{COSC.\#1} (see text for details).
Relevant datasets (see \cref{sec1:fig1}) are plotted for the output nodes of 
both the primary (dashed line plots) and secondary (solid line plots) oscillator 
nodes. In the schematic these nodes are referred to as $\text{V}_{\text{prm}}$ 
and $\text{V}_{\text{ref}}$, respectively. It is immediately observed that the calculated 
secondary oscillator ($\text{V}_{\text{ref}}$) PNOISE spectrum (\textcolor{orange}{orange solid line plot}) 
does not follow the standard Lorentzian profile (see \cref{sec1:foot5}).}
\label{sec2:fig4}
\end{figure}

\Cref{sec2:fig4}, plots the various relevant datasets 
corresponding to output nodes of both oscillators in the schematic. 
Considering the PNOISE spectra 
(orange graphs), the \texttt{PRM-OSC} graph (dashed line) 
is seen to follow the standard free-running Lorentzian characteristic 
(see \cref{sec2:fig1,sec2:fig2,sec2:fig3,sec1:foot5}). 
This makes sense as, from the above description, 
the PRM-OSC circuit must be free-running. In contrast, assuming synchronization is achieved, 
the SEC-OSC oscillator circuit must be phase-locked, 
and hence also frequency-locked, to the PRM-OSC source. 
From established ILO noise-response theory (see \emph{e.g.} \cite{Kurokawa1969,Djurhuus2025_2}) 
this implies a non-Lorentzian PNOISE spectral characteristic and 
inspecting the corresponding spectral graph in \cref{sec2:fig4} (orange solid line) 
this is indeed the case. The \texttt{QUCS-COPEN} PNOISE module, 
implemented herein, applies the rigorous COSC-PMM methodology and is hence 
is able to correctly (\emph{i.e.} without using 
any empirical/phenomenological modelling tools) 
capture the noise-response of both free-running and 
coupled circuits. The open-source \texttt{QUCS-COPEN} simulator 
project is the first ever, to our knowledge at-least, EDA program, either open-source or commercial, 
to come equipped with a noise analysis tool of this caliber.

\subsection{ Verification of the Obtained PSS Solutions \& Convergence Tests.}
\label{sec2:sub1}

\begin{table}
\centering
\includegraphics[scale=\myTableScale]{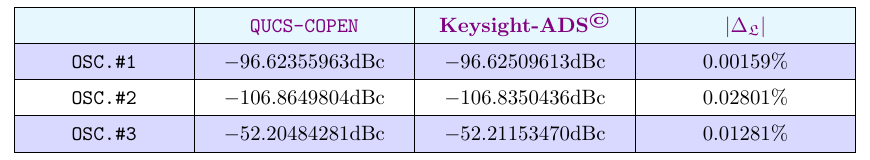}
\caption{The table lists oscillator PNOISE spectra (around 1st harmonic), evaluated @ $f_m = 1 \mathrm{kHz}$, 
calculated using the \texttt{QUCS-COPEN} package and the commercial 
Keysight-ADS\textsuperscript{\textcopyright} 
simulator for circuits : \texttt{OSC.\#1} = VDP circuit in \cref{sec2:fig1}, 
 \texttt{OSC.\#2} = BJT circuit in \cref{sec2:fig2} and \texttt{OSC.\#3} = MOSFET circuit in \cref{sec2:fig3}.  
 The relative deviation between the calculated values are 
 recorded by measure $\Delta_{\tiny \mathfrak{L}}$ (see text in \cref{sec2:sub1}).}
\label{sec2:tab1}
\end{table}

\begin{figure}[!th]
\begin{center}
\includegraphics[scale=\myFigureScaleBig]{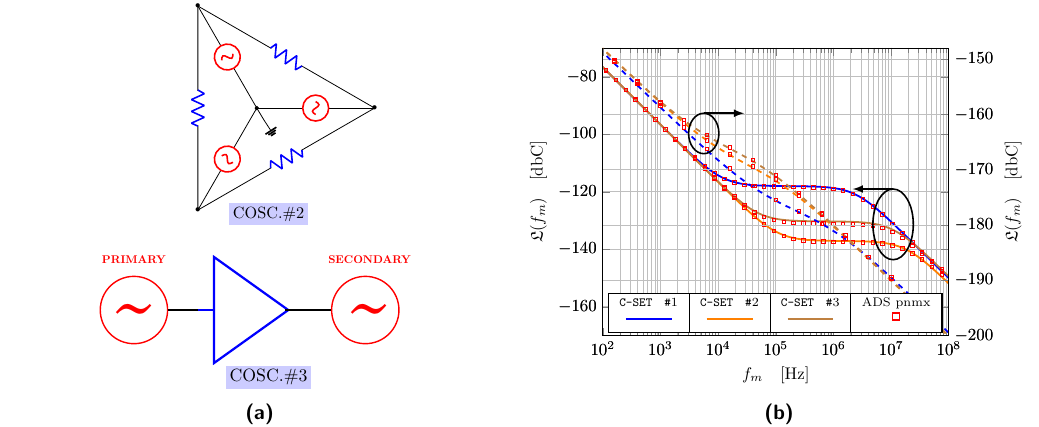}
\end{center}
\caption{PNOISE spectra of coupled oscillator circuits calculated using the 
\texttt{QUCS-COPEN} PNOISE module and with the commercial 
Keysight-ADS\textsuperscript{\textcopyright} \texttt{pnmx} routine employed as a reference. 
(a) : the two coupled circuits considered here 
are \texttt{COSC.\#2} \& \texttt{COSC.\#3}  (see text for details). 
(b) :
The PNOISE spectra are calculated for each of the coupled circuits in (a) with 
\texttt{COSC.\#2} (leftmost osc. node) drawn with \textbf{dashed line} \& 
\texttt{COSC.\#3} (secondary osc. node) drawn with \textbf{solid line}. 
The spectra are calculated 
for 3 levels of coupling strength, as specified by the parameter sets \texttt{C-SET \#1-3} (see text for details) 
with the graph color codes shown in the figure.}
\label{sec2:fig5}
\end{figure}

\newcommand\DLeqs{0.925}

Define the measure, 
$\Delta_{\mathfrak{L}} = (\mathfrak{L}^{\text{\tiny \scalebox{\DLeqs}{QC}}}(1 \mathrm{kHz}) - 
\mathfrak{L}^{\text{\tiny\scalebox{\DLeqs}{K-ADS}}}(1 \mathrm{kHz}))
/\mathfrak{L}^{\text{\tiny \scalebox{\DLeqs}{QC}}}(1\mathrm{kHz})$. 
Here, $\mathfrak{L}^{\text{\tiny \scalebox{\DLeqs}{QC}}}(f_m)$ and 
$\mathfrak{L}^{\text{\tiny \scalebox{\DLeqs}{K-ADS}}}(f_m)$ are the 
phase-noise spectra, at offset frequency $f_m$ in a 1 $\mathrm{Hz}$ bandwidth 
measured in $\mathrm{dBc}$ (see discussion in \cref{sec2}) and evaluated using the novel 
\texttt{QUCS-COPEN} PNOISE tool (QC) and the 
commercial Keysight-ADS\textsuperscript{\textcopyright} \texttt{pnmx} simulator (K-ADS), respectively. 
Thus, $\Delta_{\text{\tiny $\mathfrak{L}$}}$ records, 
for a given oscillator circuit, the relative deviation of phase-noise 
measurements evaluated @ $f_m = 1 \mathrm{kHz}$, 
produced using these two very different EDA programs. This type 
of discrete frequency verification scheme is valid 
when considering free-running oscillators due the special form of the PNOISE 
spectra produced by these circuits 
(see \cref{sec1:eq25a,sec1:foot5,sec2:fig1,sec2:fig2,sec2:fig3}). \Cref{sec2:tab1} 
displays the calculated phase-noise values and the resulting absolute measure, 
$|\Delta_{\mathfrak{L}}|$, 
for each of the three oscillators discussed above in \cref{sec2}. 
The maximum recorded error is $|\Delta_{\mathfrak{L}}| \sim 0.03\%$, or 
$|\Delta_{\mathfrak{L}}| \sim 300 \mathrm{ppm}$ (parts-per-million). The results listed 
in \cref{sec2:tab1} serve to validate the developed \texttt{QUCS-COPEN} PNOISE tool.
\par
\Cref{sec2:fig5} plots the PNOISE spectrum of two coupled oscillator 
configurations (\texttt{COSC.\#2} \& \texttt{COSC.\#3}), calculated using the \texttt{QUCS-COPEN} 
PNOISE module and the Keysight-ADS\textsuperscript{\textcopyright} \texttt{pnmx} algorithm, 
for 3 sets of coupling parameters denoted \texttt{C-SET \#1-3}. The block diagrams describing 
the coupling topology of these two circuits are shown in \cref{sec2:fig5}.(a). 
Here, \texttt{COSC.\#2} includes 3 independent oscillator units (3-ensemble), 
of \texttt{OSC.\#1}-type (see \cref{sec2:foot6}), coupled bilaterally through 
resistive coupling. The specific component values chosen for this circuit 
are reported in \cite[Fig. 7]{Djurhuus2022}.
The second circuit in \cref{sec2:fig5}.(a) is an ILO, similar to \texttt{COSC.\#1} 
considered in \cref{sec2:fig4}, but with the secondary oscillator 
represented by a \texttt{OSC.\#3}-type circuit (see 
\cref{sec2:foot6,sec2:fig3}). Then in \cref{sec2:fig5}.(b) the PNOISE 
spectra these two coupled circuits are plotted for each of the parameter (coupling) 
sets \texttt{C-SET \#1-3}, which are identical to the 3 parameter sets reported in \cite[Table 1]{Djurhuus2022}. 
Inspecting \cref{sec2:fig5}.(b), 
the calculated solution follows the reference produced by the 
commercial Keysight-ADS\textsuperscript{\textcopyright} 
\texttt{pnmx} algorithm for all offsets considered.

\section{Conclusion \& Future Work.}

\label{sec3}

The paper documents the development of a novel simulator analysis 
module aimed at calculating the response of noise-perturbed autonomous 
oscillating systems. The underlying algorithm is based on a state-of-the-art 
model proposed recently by the authors, 
and the solution was integrated into the newly established \texttt{QUCS-COPEN} EDA program; a 
clone of the open-source \texttt{QUCS} and \texttt{QUCS-S} projects. The developed 
\texttt{QUCS-COPEN} noise module is unique, emplying a novel, 
unified and rigorous time-domain methodology to construct a robust and efficient 
algorithm  uncorrupted by any phenomenological 
or empirical modelling techniques, and applicable to both free-running and coupled 
oscillating systems. 
Such features distinguish the implemented 
simulator from all other open-source and commercial EDA packages 
currently available, wherein noise analysis is, almost exclusively, performed 
by algorithms developed using some type of LTI or LTV theoretic approach, 
such as \emph{e.g.} a conversion-matrix method. The proposed \texttt{QUCS-COPEN} 
simulator project materially advances the state-of-the-art as it concerns 
the field of EDA noise simulation, analysis, synthesis and optimization of 
coupled and free-running oscillating circuits. This class of circuits is ubiquitous in 
modern communication and remote-sensing systems and the robust and efficient 
numerical tools, developed herein, will turn out to have significant 
impact in various areas of industrial circuit design.
The implemented solution was validated through a comparison-study with the commercial 
Keysight-ADS\textsuperscript{\textcopyright} EDA simulator program.
Future work will focus on improving the 
robustness and efficiency of the implemented \texttt{QUCS-COPEN} simulation tools.

\section*{Acknowledgment}

The authors gratefully acknowledge partial financial support by
German Research Foundation (DFG) (grant no. KR1016/17-1).

\end{document}